\documentclass[twocolumn,showpacs,preprintnumbers,amsmath,amssymb]{revtex4}
\usepackage{graphicx}
\usepackage{dcolumn}
\usepackage{bm}
\begin{document}

\newcommand{\RE}{\mbox{Re}}
\newcommand{\etal}{{\it et~al}}
\newcommand{\IM}{\mbox{Im}}

\title{Relativistic Doppler effect: universal spectra and zeptosecond pulses.}

\author{S.~Gordienko$^{1,2}$, A. Pukhov$^1$, O.~Shorokhov$^1$, and T.~Baeva$^1$}

\address{$^1$Institut f\"ur Theoretische Physik I,
Heinrich-Heine-Universit{\"a}t D\"usseldorf, D-40225, Germany \\
$^2$L.~D.~Landau Institute for Theoretical Physics, Moscow, Russia
}

\date{\today}

\begin{abstract}
{
We report on a numerical observation of the train
of zeptosecond pulses produced by reflection of a
relativistically intense femtosecond laser pulse from the
oscillating boundary of an overdense plasma because of the
Doppler effect. These pulses promise to 
become a unique experimental and technological tool since their length
is of the order of the Bohr radius and the intensity is extremely high
$\propto 10^{19}$~W/cm$^2$. We present the physical mechanism,
analytical theory, and direct particle-in-cell simulations.
We show that the harmonic spectrum is universal: the intensity of
$n$th harmonic scales as $1/n^{p}$ for $n < 4\gamma^2$, where
$\gamma$ is the largest $\gamma$--factor of the electron fluid
boundary, $p=3$ and $p=5/2$ for the broadband and quasimonochromatic laser
pulses respectively.
}
\end{abstract}

\pacs{03.30+p, 03.50.De, 42.65.Re, 42.65.Ky}

\maketitle

Recent impressive progress in the physics of attosecond X-ray pulses
\cite{AttoKrausz} triggers a fascinating question whether 
a range of even shorter pulses is achievable with the 
contemporary or just coming experimental technology. 
In the present work, we show that when an ultra-relativistically
intense laser pulse, $I\lambda_0^2\ge 10^{20}$W/cm$^{-2}$ $\mu m^2$,
is incident onto a  plasma-vacuum boundary, the reflected radiation
naturally contains subattosecond, or zeptosecond (zepto$=10^{-21}$)
pulses.

The idea of harmonic generation via reflection from a laser-driven
oscillating plasma boundary was first proposed a few years ago and has
been studied in several theoretical articles 
\cite{Bulanov,LewensteinMirror,Naumova}. Numerous experimental works 
\cite{Kohl,Watts,Teubner} and simulations
\cite{Gibbon,Wilks1} were devoted to the laser interactions with
overdense plasmas and the high-harmonic generation. Plaja
\etal    \cite{LewensteinMirror} were the first to realize that
the simple concept of an oscillating plasma mirror gives an
opportunity to produce extremely short pulses. In a recent work, 
Naumova \etal \cite{Naumova} have 
made a revolutionary step in this direction 
since they have proven that a single attosecond pulse can be
isolated when a laser pulse focused down to the $\lambda^3$ volume is
reflected from a plasma surface. 
 
The emphasis of our work is different. We concentrate on the very
basic physics of the high harmonic generation and discover the
spectrum universality as well as its scaling. These features have been
overseen in the earlier publications. 
The central result of our present work is the universal power-law
harmonics spectrum decaying as $1/n^p$. The exponent $p=5/2$ for a
quasi-monocromatic and $p=3$ for a broad band incident laser pulse.
 This power-law spectrum runs up to a critical harmonic number $n_c
\propto 4\gamma_{\max}^2$, where an exponential cut-off sets in.
Here $\gamma_{\max}$ is the largest relativistic factor of the plasma
boundary. It is the slow power-law decay, that allows for the
subattosecond pulses.  We compare the analytic theory with direct
particle-in-cell (PIC) simulations.

Let us consider a monochromatic laser wave with the dimensionless
vector potential $e{\bf A}_i(t,x)/mc^2=\RE \left[{\bf a}_0
\exp(-i\omega t - i\omega x/c)\right]$.  This wave is reflected by a
sharp plasma surface  
positioned at $X(t')$ at the time $t'$. First of all, we must decide 
what boundary conditions should be used. We are interested in
reflection from the overdense plasma $N_e \gg N_c$ with
$\Gamma=a_0\left(N_c /N_e \right) \ll 1$, where $N_e$ is the
plasma electron density, and $N_c = \omega^2 m_e/4\pi e^2$ is the
critical density.

The ``ideal mirror'' boundary condition implies zero tangential
components of the vector potential at the mirror surface. As a
consequence, when the ideal mirror moves with $\gamma\gg 1$
toward a laser pulse with the electric field 
$E_l$ and duration $\tau$, then the reflected pulse acquires the
electric field $E_{\mbox{\small refl}}\propto\gamma^2E_l$ and the duration 
$\tau_{\mbox{\small refl}}\propto\tau/\gamma^2$. Consequently, the 
energy of the reflected pulse must be $\gamma^2$ times higher than
that of the incident one. However, as the plasma surface is driven by
the same laser pulse, this scaling is energetically prohibited, and
the plasma cannot serve as an ``ideal mirror''. Indeed, the ideal
mirror must support a surface current $J_m \propto e N_c
\gamma a c^2 / \omega$ growing with the $\gamma-$factor. A realistic
plasma surface does not provide such a current and the boundary
condition must be changed.

Let us consider the tangential vector potential component of a laser
pulse normally incident onto an overdense 
plasma slab:

\begin{equation}
\label{Maxwell}
\frac{1}{c^2}\frac{\partial^2{\bf A}(t,x)}{\partial
t^2}-\frac{\partial^2{\bf A}(t,x)}{\partial x^2}=\frac{4\pi}{c}{\bf
j}(t,x), 
\end{equation}

\noindent where ${\bf A}(t,x=-\infty)=0$ and ${\bf j}$ is the
tangential component of plasma current 
density. Eq. (\ref{Maxwell}) yields

\begin{equation}
\label{Green}
{\bf
A}(t,x)={2\pi}\int\limits_{-\infty}^{+\infty}{\bf J}\left(t,x,t',x'\right)\,dt'dx'.
\end{equation}

\noindent Here ${\bf J}\left(t,x,t',x'\right)={\bf
j}\left(t',x'\right) \left(\Theta_--\Theta_+\right)$. We have defined
$\Theta_-=\Theta\left(t-t'-|x-x'|/c\right)$ and
$\Theta_+=\Theta\left(t-t'+(x-x')/c\right)$, $\Theta(t)$ is the
Heaviside step-function. 
Due to this choice of ${\bf J}$ the vector potential ${\bf
A}(t,x)$ satisfies both Eq. (\ref{Maxwell}) and the boundary condition
at $x=-\infty$ since ${\bf J}(t,x=-\infty,t',x')=0$. 
The tangential electric field is ${\bf E}_t =
-(1/c)\partial_t {\bf A}(t,x)$. At the electron fluid surface $X(t)$
we have

\begin{eqnarray}
\label{Integrated}
{\bf E}_t(t,X(t))=
\frac{2\pi}{c}\sum\limits_{\alpha=-1}^{\alpha=+1}\alpha\int\limits_{0}^{-\infty}{\bf j}(t+\alpha\xi/c,X(t)+\xi)\,d\xi. 
\end{eqnarray}

\noindent where $\xi = x' - X(t)$. 
If the characteristic time $\tau$ of the 
skin layer evolution is long in the sense $c\tau\gg \delta$,
where $\delta$ is the plasma skin length, then we can Taylor-expand: 
${\bf j}\left(t\pm\xi/{c},x'=X(t)+\xi\right) \approx
{\bf j}(t,x')\pm \epsilon$, where $\epsilon = (\xi/c)
\partial_{t}{\bf j}(t,x')$. We substitute this expression into
(\ref{Integrated}). The zero order terms cancel one another and we
get ${\bf E}_t(t,X(t))\propto J_p (\delta/c\tau) \ll E_l$, where
$J_p\propto cE_l$ is the maximum plasma surface 
current. Thus, as long as the skin-layer is thin and the plasma 
surface current is limited, we can use the Leontovich boundary
condition \cite{Landau8}

\begin{equation}
\label{boundaryCondition}
{\bf E}_t(t,X(t))=0.
\end{equation}

\noindent
The same boundary condition was postulated {\it ad hoc} by 
Plaja \etal  \cite{LewensteinMirror} to interpret their PIC
simulation results. In the present work, we substantiate the boundary
condition  (\ref{boundaryCondition}) physically. Using it
we are able to derive very general properties of the reflected radiation.

\begin{figure}
\centerline{\includegraphics[width=8.6cm,clip]{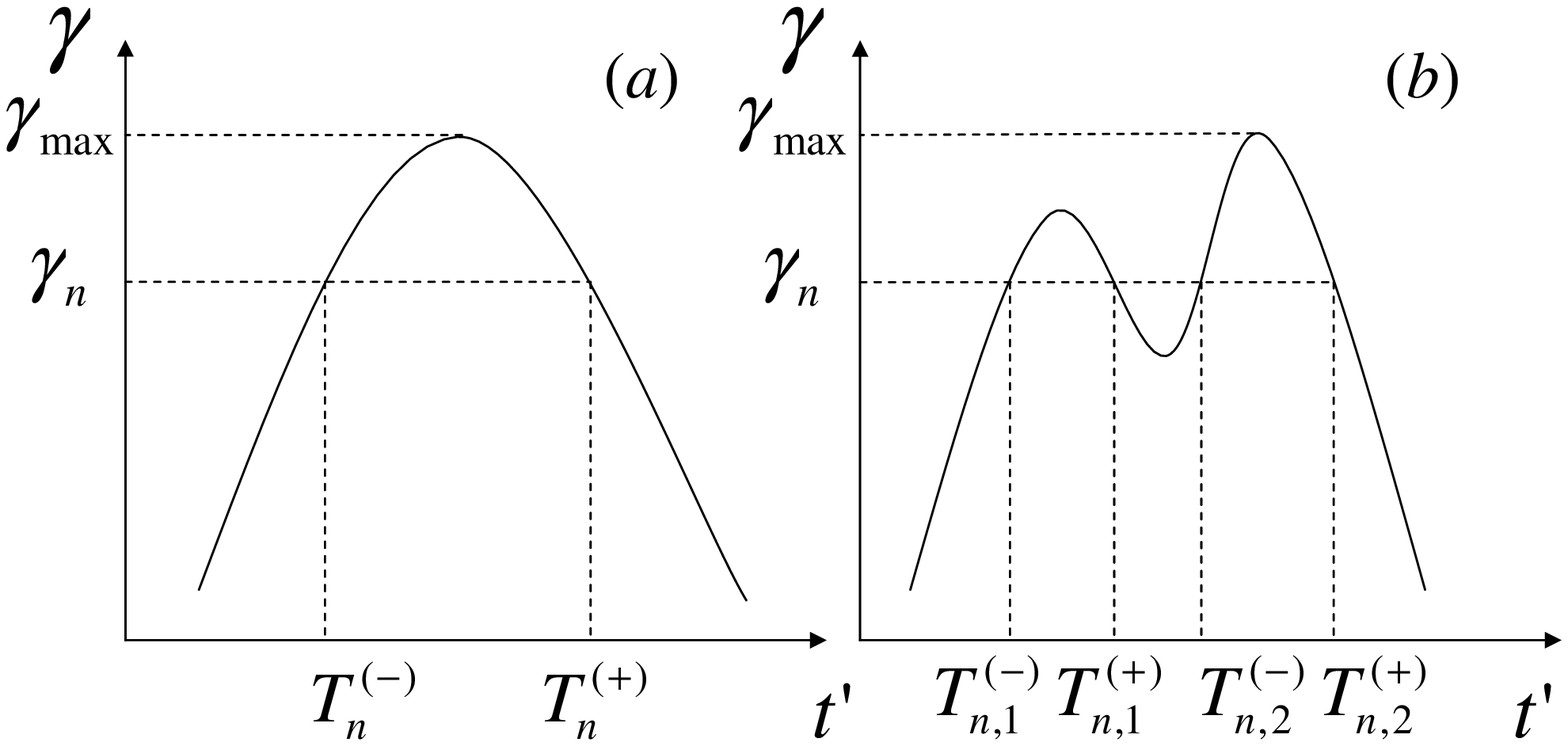}}
\caption{
A sketch of possible motions of the plasma surface. (a) Single
``hump'' per half-laser period: two saddle points are responsible for the 
generation of $n$th harmonic. (b) More complex plasma surface motion, 
the saddle points appear in pairs, two per ``hump''. Interference
between the different saddle points leads to
modulations of the spectrum \cite{Watts}.} 
\label{SaddlePoints} 
\end{figure}

According to Eq. (\ref{boundaryCondition}), the reflected wave electric
field at the plasma surface is ${\bf E}_r(t',X(t'))=-{\bf
E}_i(t',X(t'))$, where ${\bf E}_i(t', X(t'))=-(1/c)\partial_{t'}{\bf
A}_i(t',X(t'))$ is the incident laser field, $t'$ is the reflection time.
The one-dimensional (1d) wave equation simply translates a
signal in vacuum. Thus the reflected wave field at the observer
position $x$ and time $t$ is ${\bf E}_r(t,x)=-{\bf E}_i(t',X(t'))$.
Setting $x=0$ at the observer position we find that
the Fourier spectrum of the electric field ${\bf E}_r(t,x=0)$ coincides
with the spectrum of $F(t)=(A_0\omega/c)\cos(2\omega t'-\omega t)$, 
where

\begin{equation}
\label{retardation}
t'-X(t')/c=t.
\end{equation} 

\noindent is the retardation relation.
The fine structure of the spectrum of $F(t)$ depends on
a particular surface motion $X(t)$, which is defined by complex
laser--plasma interactions at the plasma surface.  The previous
theoretical works on high order harmonic generation from plasma
surfaces \cite{Bulanov,LewensteinMirror} tried to
approximate 
the function $X(t)$.

It appears, however, that a universal spectrum scaling can be obtained
without an exact knowledge of the function $X(t)$. Being interested in
universal results we discard calculating $X(t)$ and this makes our
approach very different from the previous ones. We only suppose for a
moment that the boundary motion is periodic $X(t+\pi/\omega) =
X(t)$. Later we consider non-monochromatic laser pulses and get rid
even of this restriction. 
%%%%%%%%%%%%%%%%%%%%%
% using the Fourier integral enables us to get rid of even this
% restriction. 
%Only non-monotonous dependance of the surface $\gamma$-factor of time
%is %important for the overall spectrum scalings.  
% mi  predpolagaem periodichnost', chtobi imet' tol'ko nechetnie
% garmoniki. %Kogda %mirassmatrivaem korotkij impul's I vmesto rjada
% Fourier pishem integral %Fourier 
%nam dazhe eto ne nado
%%%%%%%%%%%%%%%
First, we mention that the Fourier spectrum of $F(t)$ can be
represented in the form:

\begin{eqnarray}
\hat{F}_n = {A_0\omega}
\left[\hat{C}_{n+1} + \hat{C}_{n-1} - i(\hat{S}_{n+1} - \hat{S}_{n-1})
\right]/(2c), \label{Fourier}
\end{eqnarray}

\noindent where $\hat{C}_n$ and $\hat{S}_n$ are the $n-$th harmonics
of $C(t) = \cos \left(2\omega t'\right)$ and  $S(t) = \sin\left(2\omega t'\right)$,
and $t'$ is the retarded time (\ref{retardation}). 
 We examine only the spectrum of $C(t)$, because that
of $S(t)$ can be worked out analogously. 

It is easy to see that the function $C(t)$ has a period
$\pi/\omega$. Thus, its spectrum contains only even laser harmonics.
Introducing the new variable $\tau=2t$ we obtain

\begin{equation}
\label{C_{2n}1}
\hat{C}_{n=2m}=\frac{1}{2}\int\limits_{-\pi/2\omega}^{\pi/2\omega}
\left[\exp\left(i\Phi_1(\tau)\right)+\exp\left(i\Phi_2(\tau)\right)\right]\,d\tau, 
\end{equation}

\noindent where $\Phi_1(\tau)=(1-m)\omega\tau+2\Phi_r(\tau)$,
 $\Phi_2(\tau)=-(1+m)\omega\tau-2\Phi_r(\tau)$, and
%\begin{equation}
%\label{retPhase}
$\Phi_r(\tau)=({\omega}/{c}) X\left(
\arccos C\left({\tau}/{2}\right)/(2\omega)\right)
$%\end{equation}
%\noindent 
 is the retarded phase. The definition of $\Phi_r(\tau)$ is recurrent,
because $C(t)$ itself is defined through $X$.

\begin{figure}
\centerline{\includegraphics[width=8.6cm,clip]{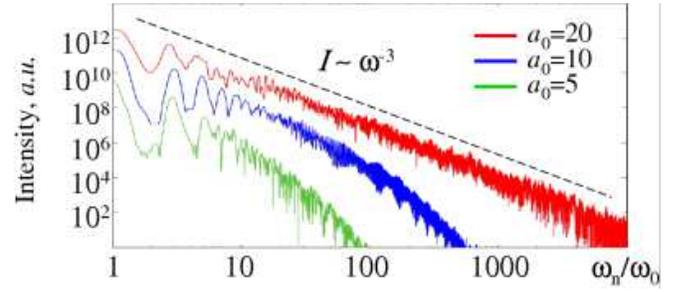}}
\caption{Spectra of the reflected radiation for the laser amplitudes
$a_0=5,10,20$. The broken line marks the universal scaling $I\propto
\omega^{-3}$.} \label{figSpectra} 
\end{figure}

To calculate the spectrum (\ref{C_{2n}1}) we use the saddle
point technique. The saddle points $\tau_n$ are obtained from the equations
$d\Phi_1(\tau_n)/d\tau=0$ and $d\Phi_2(\tau_n)/d\tau=0$. The first
equation reduces to $d\Phi_r(\tau_n)/d\tau = n/2-1$. Taking the first
derivative of the recurrent definition of $\Phi_r$ we re-write the
saddle-point equation as

\begin{equation}
\label{saddle}
dX(T_n)/dT=c(1-2/n)~~\mbox{or}~~n \approx 4\gamma^2(T_n), 
\end{equation}

\noindent where $T(\tau)=(1/2\omega)\arccos C(\tau/2)$,
$T_n=T(\tau_n)$ and $\gamma=1/\sqrt{1 - \beta^2}$, $c\beta =
dX(T)/dT$. 
Eq.~(\ref{saddle}) has a clear physical meaning. The reflected radiation
frequency is multiplied by the factor $4\gamma^2$ because of the relativistic Doppler effect, where $\gamma$ is
the relativistic factor of the plasma surface.
If the plasma surface oscillates non-relativistically, so that
$\gamma \approx 1$, then Eq. (\ref{saddle})
has no real solutions for $n>1$, and the spectrum of $C(t)$
exponentially decays. When $\gamma \gg 1$, there is a real
solution for any $n<n_{c}=4\gamma_{\max}^2$. A similar consideration
shows that the second saddle-point equation has no real solutions.
The spectrum is thus

\begin{equation}
\hat{C}_{n} = \sum_{\tau_n}\sqrt{{2\pi}/{|d^2_t\Phi_r(\tau_n)|}}
\exp\left(i\Phi_1(\tau_n) \pm
i{\pi}/{4}\right).
\label{spectrum}
\end{equation}

\noindent To estimate (\ref{spectrum}) we use the fact that the
highest harmonics are generated 
around the time $T_{\max}$, when the plasma surface moves toward the
laser with the highest relativistic factor $\gamma_{\max}$. In its
vicinity, one can approximate $\gamma \approx \gamma_{\max}[1 -
g^2(T-T_{\max})^2/2]$ as shown in Fig.~\ref{SaddlePoints}a. A
straightforward algebra leads to 

\begin{equation}
\label{theSecondDerivative}
\left|{d^2_{t}}\Phi_r(\tau_n)\right| ={D g\omega n^2}\sqrt{{n(n_c-n)}/{n_c}}+O(n^2),  
\end{equation}

\noindent where  $D=\sin(2\omega T_{max})/2\gamma_{max}$. The estimate
(\ref{theSecondDerivative}) leads to the spectrum intensity scaling

\begin{equation}
\label{spectr52}
\hat{C}_{n}^2 \propto n^{-5/2}~~~\mbox{for}~~~n<n_c. 
\end{equation} 

When one considers the physical mechanism of high-harmonic generation
as presented in Fig.~\ref{SaddlePoints}, it becomes evident that the
harmonics are emitted in the form of 
subattosecond pulses. Indeed, all harmonics above a number $n<n_c$ 
are generated at times between $T_{n}^{(-)}$ and
$T_{n}^{(+)}$. From Eqs. (\ref{retardation}), (\ref{saddle})
the subattosecond pulse duration $\Delta \tau=t(T_{n}^{(+)})-t(T_{n}^{(-)})$ as it is seen by the observer
is $\Delta \tau =({4}/{3gn})\sqrt{{n_c}/{n}}$ for $1\ll n\ll n_c$. This estimation tells us
that the reflected pulse can be made very short by applying a filter
selecting harmonics with high numbers larger than $n$.

\begin{figure}
\centerline{\includegraphics[width=8.6cm,bb=15 454 580 813, clip]{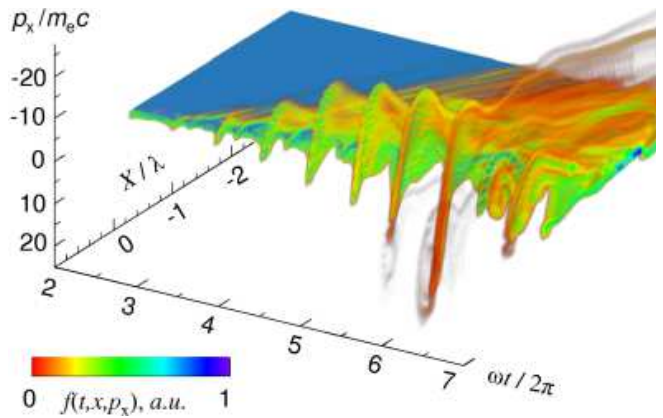}}
\caption{Electron distribution function
The helix represents
the electron surface motion in the laser field. The reddish
downward spikes stay for the
surface relativistic motion towards the laser. These spikes are responsible for
the zeptosecond pulse generation.} \label{fig3d} 
\end{figure}
 
The mechanism presented in Fig.~\ref{SaddlePoints} has another very
interesting consequence. Each harmonic is generated due to the saddle points corresponding to the proper $\gamma$--factor. These saddle points
come into (\ref{spectrum}) with different phase
multipliers. Fig.~\ref{SaddlePoints}b represents the case of a more
complicated plasma surface motion, when the $\gamma-$ factor has
several maxima, ``humps'', per half-laser period. Yet, one can see that the saddle
points are grouped, i.e. there is a couple of the saddle points on
every "hump".  The phase shift between the contributions of the saddle
points $T_{n,i}^{(\pm)}$ belonging to the the same $i$th "hump" is not
very large. For example, for the saddle points $T_n^{(\pm)}$ this
phase shift equals  
$
\Phi(n)=-({5\omega}/{3g})\sqrt{{n_c}/{n}}
$. The frequency modulation period due to their interference is 
$\hat{\Omega}\propto 2\pi\omega/(d\Phi(n)/dn)$,
i.e. $\hat{\Omega}\propto (6\pi g/5\sqrt{n_c})n^{3/2}$. This
inteference really brings modulation to the spectrum only if
$\sqrt{n_c}\gg 6\pi g/(5\omega)$. On the other side, the phase shift between the
contributions from different "humps" can be much larger. This means
that a non--trivial motion of the critical surface producing more than
one $\gamma$--factor "hump" per oscillation period is the cause
of the spectrum modulation. This conclusion supports the explanation
of the modulation proposed by Watts \etal  \cite{Watts} and
agrees with the experimental observations by Teubner \etal
 \cite{Teubner} and Watts \etal  \cite{Watts}. Finally, we 
notice that the larger number of the saddle points does not change the
averaged value for $d^{2}_t\Phi(t_n)$ and, consequently, does not affect the overall
spectrum scaling.

\begin{figure}
\centerline{\includegraphics[width=8.6cm,clip]{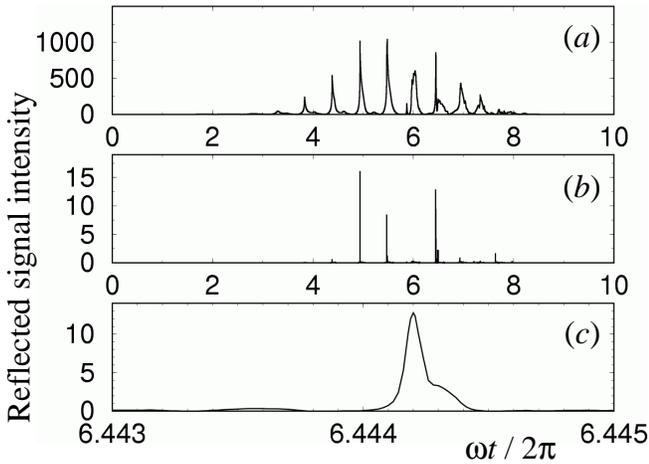}}
\caption{Zeptosecond pulse train: a) temporal structure of the
reflected radiation; b) zeptosecond pulse train seen after spectral
filtering; c) one of the zeptosecond pulses zoomed, its 
FWHM duration is about 300~zs.} \label{figZepto}  
\end{figure}

A careful analysis reveals that the intensity spectrum $\propto
1/n^{5/2}$ is valid for a monochromatic incident wave only. Indeed, if
the laser pulse is short and has a spectral bandwidth $\Delta \omega$,
then the spectral scaling fails at frequencies $\Omega >
\omega^2/\Delta\omega$. At these frequencies, the reflected radiation
is broadband and contains no distinguishable harmonics.
To examine the finite bandwidth influence we represent the incident
wave as ${\bf E}_i(t,x)=\RE\left[\omega {\bf A}_0\exp(-i\phi(t +
x/c))\right]$, where  $\partial_t\phi(t+ x/c)\approx \omega$ and
$\Delta\omega\ll\omega$. A procedure almost identical to that 
leading to Eqs. (\ref{C_{2n}1}) and (\ref{spectrum}) reduces the
Fourier spectrum calculation of ${\bf E}_r$ to that of $C(t)=\exp 
\left(-i\phi\left(T(t)\right)\right)$, where $T(t)=-t+2X(t')/c$,
$2t'=t+F\left(i\ln C(t)\right)$ and $F(\phi(t))=t$ for any $t$. 

To apply the steepest descent method to evaluate the $\Omega$-Fourier
amplitude $\hat{C}_{\Omega}$ we introduce a new function
$\Phi=\RE(\phi)$ such that $d_t\Phi(-t+2X(t')/c)=\Omega$. We have
neglected $\IM(\phi)$ because 
 the shift of the saddle point due to $\IM(\phi)$ is only
$O((\omega/\Omega)^2)$. Instead of the monochromatic expression
(\ref{spectrum}) we obtain now  $\hat{C}_{\Omega}\propto
1/\sqrt{\left|d^2_t\Phi\right|}$, where 

\begin{equation}
\label{derivatives}
d^2_t\Phi=\left(\frac{\Omega}{\omega}\right)^3\left(B+O\left(\frac{g\omega}{\gamma_{max}}\sqrt{\frac{(2\omega
n_{c}-\Omega)}{n_{c}\Omega}}\right)\right) 
\end{equation}

\noindent for $\Omega<2\omega n_{c}$. Here, the $O(\dots)$-term on the
right hand part of Eq.(\ref{derivatives}) is inherited from the
monochromatic approximation, see Eq. (\ref{theSecondDerivative}). The
new term
$B=\left|{2\omega
C_{max}{d}/{dT}\left[({d\Phi(T_{max})/dT})/({d\phi(T_{max})/dT})\right]}\right|$ 
appears due to the finite spectral width. Here  $T_{max}$, $n_{c}$ and
$g$ are the same as in Eq. (\ref{theSecondDerivative}), and
$C_{max}=C(T_{max})$. 

One readily finds that $B\propto 1/\tau^2$, where 
$\tau\propto1/\Delta\omega$ is the incident pulse duration. Thus,
we can neglect the $O(\dots)$-term for $\Omega/\omega > n_{c1} = (g\omega)/(\gamma_{max}^2
B^2)$ what leads to 

\begin{equation}
\label{mainResult}
I_\Omega\propto \left|\hat{C}_{\Omega}\right|^2 \propto
1/\Omega^3,~~~~~~~\Omega = n\omega 
\end{equation}

\noindent We emphasize that the criterion $\Omega>n_{c1}$ defining the transition from $1/\omega^{5/2}$-spectrum to
$1/\omega^3$-spectrum depends on the function $B$ that will be analised in detail elsewhere. 
Altogether, the spectral intensity $I_n$ of the reflected
radiation scales as  $I_\Omega\propto1/n^{5/2}$ for $1 < n < n_{c1}$, 
then as $I_\Omega \propto 1/n^3$ for $n_{c1}<n<4\gamma_{\max}^2$ and finally decays
exponentially for $n>4\gamma_{\max}^2$.

To check our analytical results, we have done a number of 1d PIC
simulations. A laser pulse with the Gaussian profile
$a=a_0\exp\left(-t^2/\tau_L^2\right)$ was incident onto a plasma layer with a
step density profile. Fig.~\ref{figSpectra} shows spectra of the
reflected radiation for laser amplitudes $a_0=5,10,20$, duration
$\omega \tau_L = 4\pi$ and the plasma density $N_e = 30 N_c$, which 
roughly corresponds to the solid hydrogen or liquid helium. 
The
log-log scale of  Fig.~\ref{figSpectra} reveals the power-law scaling
of the spectral intensity $I_\Omega \propto 1/n^3$. The critical
harmonic number $n_c$, where the power-law scaling changes into the
exponential decay increases for higher laser amplitudes. Also, the
spectral intensity modulations are seen
\cite{Teubner,Watts}. 

Let us take a close look at the particular case $a_0=20$ (the red line in Fig.~\ref{figSpectra}). In this
case, the power-law spectrum extends above 
the harmonic number
2000, and zeptosecond pulses can be generated. As one sees from the
electron distribution function $f(t,x,p_x)$, Fig.~\ref{fig3d}, the
maximum surface $\gamma-$factor $\gamma_{\max}\approx
25$ is achieved at the time $t\approx 6$. The temporal profile of the
reflected radiation is shown in 
Fig.~\ref{figZepto}. When no spectral filter is applied,
Fig.~\ref{figZepto}a, a train of attosecond pulses is observed
\cite{LewensteinMirror}. However, when we apply a spectral filter selecting
harmonics above $n=300$, a train of much shorter pulses is recovered,
Fig.~\ref{figZepto}b. Fig.~\ref{figZepto}c zooms to one of these
pulses. Its full width at half maximum is about $300~$zs. At the same
time its intensity normalized to the laser frequency is huge
$(eE_{zs}/mc\omega)^2 \approx 14$ that would 
correspond to the intensity $I_{zs}\approx 2\times10^{19}$~W/cm$^2$.

The presented theory and simulations are 1d, i.e., in a planar
geometry. This assumes that the laser focal spot radius
$r\gg\lambda$. Recent 2d and 3d PIC simulations in the $\lambda^3$
regime \cite{Naumova} have shown that the multi-dimensional effects
may help, particularly, to isolate a single (sub-)attosecond pulse.

In conclusion, the reflection from ultrarelativistically moving plasma
boundaries forms two different universal spectra because of the
relativistic Doppler effect: $1/n^{5/2}$ for an incident
quasimonochromatic wave and $1/n^3$  for a broadband pulse. The slow
power law decay of the high harmonic spectrum causes 
high intensity zeptosecond pulses. The observed pulse duration of
300~zs corresponds to the spatial pulse extent comparable to the first
Bohr orbit. As a result, the special relativity
theory introdicing the relativistic Doppler effect may open a new page
of applied ultrafast laser spectroscopy.

We gratefully acknowledge  discussions with
Prof. J.~Meyer-ter-Vehn and Dr. G.~D.~Tsakiris. The work has been
supported by RFFI 04-02-16972, the AvH fund and DFG.

\end{document}